Original Research Paper

# Factors Influencing Intention to use the COVID-19 Contact Tracing Application

**¹Vinh T. Nguyen and ²Chuyen T. H. Nguyen**

*¹Department of Information Technology, TNU-University of Information and Communication Technology, Vietnam*
*²Department of Primary Education, Thai Nguyen University of Education, Vietnam*

Corresponding Author:
Vinh T. Nguyen
Department of Information
Technology, TNU-University
of Information and
Communication Technology,
Vietnam
Email: vinhnt@ictu.edu.vn

**Abstract:** Due to the emergence of the COVID-19 pandemic, governments have implemented several urgent steps to minimize the disease's effect and transmission. Supportive measures to trace contacts and warn people infected with COVID-19 were also implemented such as the COVID-19 contact tracing application. This study investigated the effects of variables influencing the intention to use the COVID-19 tracker. The extended Unified Theory of Acceptance and Use of Technology model was used to investigate user behavior using the COVID-19 tracker application. Google Form was used to construct and distribute the online survey to participants. Experiment results from 224 individuals revealed that performance expectations, trust, and privacy all have an impact on app usage intention. However, social impact, effort expectation, and facilitating conditions were not shown to be statistically significant. The conceptual model explained 60.07% of the amount of variation, suggesting that software developers, service providers, and policymakers should consider performance expectations, trust, and privacy as viable factors to encourage citizens to use the app. This study work's recommendations and limitations are thoroughly discussed.

## Introduction

The outbreak of the COVID-19 pandemic at the beginning of 2020 and the recent significant epidemic have had a severe impact on all countries worldwide (Sudheer Reddy *et al*., 2020; Mahmoud *et al*., 2020; Nguyen, 2022). As of mid-October 2021, there were around 240 million infections worldwide, with 4.86 million fatalities (Dong *et al*., 2020). Faced with this dilemma, governments throughout the world have implemented several urgent steps to minimize the disease's effect and transmission, including travel restrictions, social isolation, mask use, and the closure of public areas. In tandem with initiatives to raise public knowledge of epidemic prevention through media such as television, newspapers, radio, social networks, and text messaging (Binsar and Mauritsius, 2020), etc., governments have also implemented many supportive measures to trace contacts and warn people infected with COVID-19 (Grekousis and Liu, 2021). For example, the COVID-19 self-reported symptom and contact tracking apps are the two types of apps widely being used. The former allows people to report their health conditions and the latter enables them to check and trace their contacts. Both provide additional information such as the symptoms of COVID-19, updates on COVID-19, the rate at which the virus spreads in various locations, high risky areas in the country, etc. This kind of app is anticipated to assist state authorities in swiftly tracking and managing infections in the community and users will be able to collect timely information to help avoid epidemics (Bansal *et al*., 2010; Wang *et al*., 2020). To interact with devices, the contact tracking app essentially uses Bluetooth low-energy technology. In this case, your phone will register any other phones it comes into contact with, as long as both your device and the others have a fully enabled COVID-19 contact tracking app. They use random numerical ID numbers that change regularly and are destroyed after 14 days (COVID-19's incubation period) (Fetzer and Graeber, 2021).

Even though the COVID-19 app is supposed to have a significant positive impact and that many people would utilize it, the actual usage was not as expected. For example, Hargittai *et al*. (2020) reported that only 67% of respondents were willing to install a tracking app for several reasons, the majority of which are privacy concerns. Similarly, Garrett *et al*. (2021) estimated that the acceptability of tracking technologies ranges from 62-to 70% depending on the scenario. Many more other examples can be found in the literature (O'Connell *et al*., 2021; Grekousis and Liu, 2021). This issue raises a research question: What factors influence the use of a COVID-19 tracking application?

Answering the above question is critical to persuading individuals to participate in and support epidemic prevention in both physical and virtual spaces. The findings will help managers, policy researchers and software developers make the required modifications to promote the benefits and power of the COVID-19 tracking app. As Colizza *et al*. (2021) pointed out "Time to evaluate COVID-19 contact-tracing apps", literature work has made major contributions to this topic from multiple perspectives (e.g., qualitative and quantitative) (Grekousis and Liu, 2021; Hargittai *et al*., 2020). However, no study has attempted to explain the causal connection between factors. As a result, this research has a distinct and important position in the current landscape, especially given that the COVID-19 epidemic shows no signs of abating.

## Materials and Methods

### Conceptual Framework and Research Hypotheses

Along with the continuous development of modern digital devices, the software is constantly being developed, updated, and upgraded to help users manage and use data more effectively. However, for software to be used effectively, it requires a lot of effort and resources. Lessons learned from previous failures show that, if software development is not scrutinized, it will have huge consequences in terms of both time and money (Braude and Bernstein, 2016). To alleviate the issues, one of the promising approaches is to evaluate the applications from multiple perspectives (e.g., algorithms, techniques, user behaviors). In terms of user behaviors or user acceptance, the Technology Acceptance Model (TAM) is one of the widely used methods to anticipate the degree of acceptance of an application (Davis, 1989). The TAM model focused on four dimensions including actual system usage, behavioral intention, Perceived Ease of Use (PEU), and Perceived Usefulness (PU). The model asserts that when a new app was introduced to users, its actual usage was directly influenced by behavioral intention and indirectly impacted by PEU and PU. Over the years, TAM has been extended to include additional factors (e.g., task-technology fit, output quality, visual design, subjective norm, and result demonstrability) (Jung *et al*., 2021). The addition of additional factors has created major challenges for novice researchers not specializing in social science behavior. To alleviate the aforementioned issues, Venkatesh *et al*. (2003) unified eight prior models into the Unified Theory of Acceptance and Use of Technology (UTAUT) to express user behavior toward an application. The author asserted that user behavior can be predicted by four factors including expected performance, expected effort, facilitating conditions, and social influence. Following this approach, the current study extended the UTAUT model with two additional components: Perceived risk and trust.

### Behavioral Intention

A vast array of behavioral research and theories seeks to examine internal and external factors that have an impact on user behavior (Venkatesh *et al*., 2003; Nguyen *et al*., 2022). An individual's behavioral intention can be described as the perception of whether (s) he will engage in a certain activity (Fishbein and Ajzen, 1977). The current research defined behavioral intention as an individual's likelihood of using the COVID-19 tracking app. To measure behavioral intention, three questions were used: (1) I will continue to use the COVID-19 tracker for the next 6 months, (2) I will still use the COVID-19 tracker every day, and (3) I will recommend the COVID-19 tracker to my friends.

### Performance Expectancy

This factor is termed as an individual's belief that utilizing an IT software would let (s)he to meet his performance goal (Venkatesh *et al*., 2003). The current research used three questions to measure performance expectancy, including: (1) Using a COVID-19 tracker allows me to grasp information about COVID more quickly, (2) Using a COVID-19 tracker helps me improve the effectiveness of COVID prevention, (3) Using COVID-19 tracker helps me to timely grasp the necessary information where I live. The following hypothesis was proposed:

Hypothesis 1 ($H_1$): Behavioral Intention was positively influenced by performance expectancy

### Effort Expectancy

In short, this factor describes how easy it is to use the system (Venkatesh *et al*., 2003). In the UTAUT model, effort expectancy is a critical predictor. In this study, effort expectation depicts users' perceptions about the ease of use of the COVID-19 tracker. To measure effort expectation, we asked four questions, which are as follows: (1) Learning how to use the COVID-19 tracker is relatively easy for me, (2) COVID-19 tracker functions and operations are clear and easy to understand, (3) COVID-19 tracker app is easy to use and (4) I easily master COVID-19 tracker app. The following hypothesis was proposed:

Hypothesis 2 (H₂): Effort expectancy positively influences behavioral intention

### Social Influence

This factor refers to the belief that an individual should use a particular software if it was recommended by influential users (Venkatesh *et al.*, 2003). Studies in UTAUT indicated that social influence had a statistically significant impact on behavioral intention since it alters the beliefs of potential users. The current study terms social influence as friends, family members, and coworkers who persuade a person to adopt new technologies. We utilized three questions to measure social impact namely: (1) My family members think that I should use the COVID-19 tracker, (2) My friends and colleagues think I should use the COVID-19 tracker, and (3) I use the COVID-19 tracker because it is advertised from the media. The following hypothesis was proposed:

Hypothesis 3 (H₃): The social influence had a positive and direct impact on behavioral intention

### Facilitating Condition

This factor refers to the belief that individuals would utilize an IT system if supportive technologies are available to them (Venkatesh *et al.*, 2003). The following questions were used to measure facilitating conditions in the current research: (1) I have a device on which to install the COVID-19 tracker (e.g., phone, tablet), (2) my devices are compatible with the COVID-19 tracker, and (3) I have support when I have problems with COVID-19 tracker. Thus, the following hypothesis was proposed:

Hypothesis 4 (H₄): Facilitating conditions had a positive impact on behavioral intention

### Trust

Trust indicates a readiness to be vulnerable in the face of favorable anticipation of future conduct from the external (Mayer *et al.*, 1995). Existing research indicated that trust influenced behavioral intention and risk perception (Alfina *et al.*, 2014; Beldad *et al.*, 2010). Alfina *et al.* (2014) anticipated that trust would be connected to performance and effort expectations. The present research utilizes three questions to measure trust: (1) I believe that the information provided by the COVID-19 tracker is reliable, (2) I trust the use of the COVID-19 tracker, and (3) COVID-19 tracker provides the functionality that users need. With the aforementioned studies, the following hypotheses were proposed:

Hypothesis 5 (H₅): Trust will have a positive effect on behavioral intention
Hypothesis 6 (H₆): Trust will have a positive effect on performance expectancy
Hypothesis 7 (H₇): Trust will have a positive effect on effort expectancy
Hypothesis 8 (H₈): Trust will harm privacy risk

Privacy Risk: Privacy risk is considered a user's anxiety about the disclosure of personal information (Li, 2011). The current study defines privacy risk as to the chance of a person incurring a loss when using the COVID-19 tracker. The less the users perceive risk, the happier they are. Many studies have found that privacy issues reduce user confidence and, as a result, the desire to utilize the system (Nguyen, 2021; Alotaibi, 2014). Two questions were used to evaluate perceived risk, namely: (1) I think using the COVID-19 tracker puts my privacy at risk, and (2) My personal data may be leaked when using the COVID-19 tracker. As such, the following hypothesis was proposed:

Hypothesis 9 (H₉): Privacy risks harmed behavioral intention

These assumptions were transformed into the study model depicted in Fig. 1, representing a causal connection schema, and served as the starting point for this investigation. The ellipses indicate the constructs (also known as latent variables) assessed by a series of items, while the arrows represent hypotheses numbered 1 to 9.

### Data Collection and Subjects of Interest

The study utilized a non-probability, purposive sampling approach to gathering data for analysis. Google Form was used to create and share the questionnaires with participants during the period from June 18, 2021, to June 25, 2021. Participants of interest were recruited to participate in the study via personal email or social networking sites (e.g., Facebook, Twitter). Target subjects are those who had experience with the COVID-19 tracker. The expected number of users that participated in the study was 400 and the response rate is 295 (73.75%). The total number of responses included in the analysis was 224 (75.93%).

There are several debates regarding the sample size in the literature, so its determination is inconsistent. Some studies asserted that an experiment should include at least 100 to 200 subjects (Kock and Hadaya, 2018). The acceptable range is usually from 300 to 500 (Kock and Hadaya, 2018), or with proportion to a free parameter.

**Table 1:** Construct and items

Performance expectancy (Venkatesh *et al*., 2003)

(PE 1) Using the COVID-19 tracker allows me to grasp information about COVID more quickly
(PE 2) Using the COVID-19 tracker helps me improve the effectiveness of COVID prevention
(PE 3) Using the COVID-19 tracker helps me to timely grasp the necessary information where I live

Effort expectancy (Venkatesh *et al*., 2003)

(EE 1) Learning how to use the Covid-19 tracker is relatively easy for me
(EE 2) COVID-19 tracker functions and operations are clear and easy to understand
(EE 3) COVID-19 tracker app is easy to use
(EE 4) I easily master the COVID-19 tracker app

Social Influence (Venkatesh *et al*., 2003)

(SI 1) My family members think that I should use the COVID-19 tracker
(SI 2) My friends and colleagues think I should use the COVID-19 tracker
(SI 3) I use COVID-19 tracker because it is advertised in the media

Facilitating conditions (Venkatesh *et al*., 2003)

(FC 1) I have a device on which to install the COVID-19 tracker (e.g., phone, tablet)
(FC 2) COVID-19 tracker is compatible with my devices
(FC 3) I have support when I have problems with the COVID-19 tracker

Privacy risk (Budi *et al*., 2021)

(T 1) I believe that the information provided by the COVID-19 tracker is reliable
(T 2) I trust the use of the COVID-19 tracker
(T 3) COVID-19 tracker provides the functionality that users need

Trust (Budi *et al*., 2021)

(T 1) I believe that the information provided by the COVID-19 tracker is reliable
(T 2) I trust the use of the Covid-19 tracker
(T 3) COVID-19 tracker provides the functionality that users need

Behavior Intention (Venkatesh *et al*., 2003)

(BI 1) I will continue to use the COVID-19 tracker for the next 6 months
(BI 2) I will still use the COVID-19 tracker every day
(BI 3) I will recommend the COVID-19 tracker to my friends

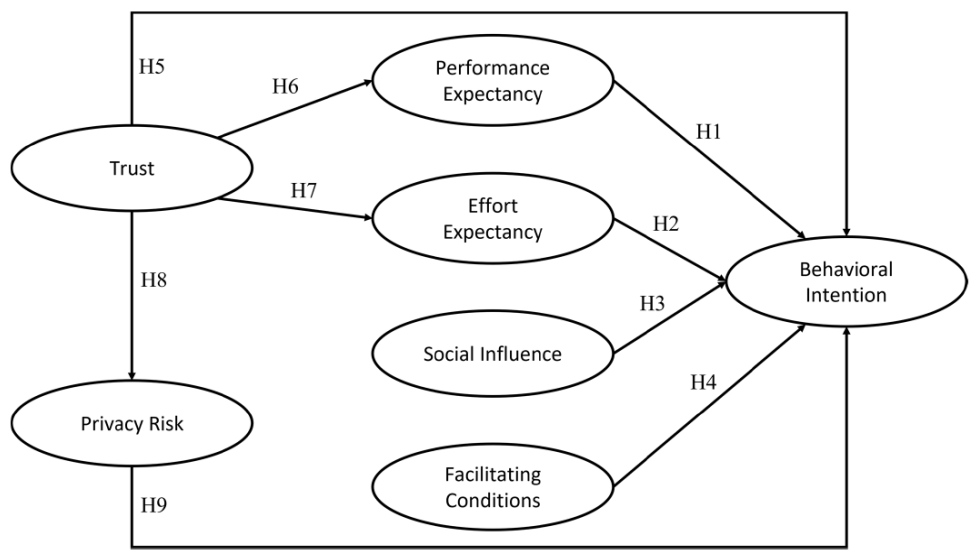

**Fig. 1:** The conceptual model depicts identified factors in the current research

Anderson and Gerbing (1984) stated that a sample of size 100 would be sufficient for convergence when there are more than three indicators and a sample size of 150 is usually sufficient for a convergent and accurate solution. The current study followed the guideline provided by Kline (2015) to assess an acceptable sample size (Soper, 2016). The configuration of the tool is set as follows: The expected effect size is 0.3, the desired statistical power level is 0.8, the number of latent variables is 7, the number of observable variables is 21 and the probability level is 0.05. As a result, the recommended minimum sample size was 200. Because the actual sample size of this research was 224, which was higher than the above-mentioned limits (200), the needed sample size for the current study was satisfied.

*Measures*

Following a review of the survey questions based on the research methodology, 21 questions were chosen and included in the study (Table 1). For each question, a five-point Likert scale (1 = Strongly Disagree, 2 = Disagree, 3 = Neutral, 4 = Agree, 5 = Strongly Agree) was utilized.

*Data Analysis*

Due to its ability to deal with a small sample size while needing strict normal distribution, Generalized Structured Component Analysis (GSCA) was used to analyze this proposed research model (Hwang and Takane, 2014). The GSCA is a structural equation model-based component that may be used to simulate Partial Least Squares (PLS) paths. Hwang and Takane (2014) introduced this approach to optimize a global function. GSCA gives a global criterion of the least square parameters, which is consistently minimized to compute model parameters, in contrast to PLS. As a result, GSCA has an overall measure of model fit while retaining all of the benefits of PLS. Furthermore, as compared to PLS, the GSCA handles more varied path analyses. This study used GSCA software (Hwang *et al*., 2019) for parameter estimations.

# Results

## Demographic Characteristics

Table 2 presents the survey data, with males accounting for 16.07% of the total, while women accounted for 83.92%. More than half of the poll respondents are students aged 10 to 20 years (52.68%), 27.23% are between the ages of 21-30, and 11.16% are between the ages of 31-40 and 8.93% are above 41 years old. COVID-19 contract tracing application users are mostly located in rural regions and mountainous areas (52.23%), with the remainder residing in metropolitan areas (28.57%) and towns (19.20%). The survey results are also compatible with the regional features of the study site, which is a hilly area.

## Quantitative Analysis

Table 3 displayed the descriptive statistics for the construct items. Items in the table indicated that all means of the extended UTAUT measures were greater than the mid point of 2.5, with standard deviations ranging from 0.6880 to 1.3707.

The internal consistency and convergent validity measurements for each concept were shown in Table 4. Dillon-Goldstein's rho was used to justify the internal consistency and reliability requirements of each construct.

Almost all of the values, which ranged from 0.7394 to 0.9194, were more than 0.7, above the acceptable reliability estimate (Hwang and Takane, 2014). We also looked at each latent variable's Average Variance Extracted (AVE) value to determine if it was convergent. All AVE values were more than 0.5, ranging from 0.7453 to 0.9470, suggesting convergent validity.

Table 5 showed the item loading estimates, as well as their Standard Errors (SEs) and 95% bootstrap percentile Confidence Intervals (CIs) with Lower and Upper Bounds (LB and UB respectively). 100 bootstrap samples were used to calculate the Confidence Intervals (CIs). At the 0.05 alpha level, a parameter estimate was considered statistically significant if the 95 percent CI did not include the value zero. The loading estimates were all statistically significant, indicating that all of the items were reliable predictors of the constructs.

Table 6 showed that GSCA provided FIT = 0.6007 (SE = 0.014, 95% CI = 0.5768-0.6291), AFIT = 0.5965 (SE = 0.0142, 95% CI = 0.5724-0.6252), GFI = 0.993 (SE = 0.0009, 95% CI = 0.9916-0.9949) and SRMR = 0.2419 (SE = 0.0194, 95% CI = 0.2171-0.2852). Both FIT and Adjusted FIT (AFIT) were used to investigate the variance in data explained by a given model configuration. FIT values range from 0 to 1. The characteristics and meanings of FIT and AFIT are equivalent to $R^2$ and adjusted $R^2$ in the linear regression. Experiment results of FIT and AFIT showed that the model accounted for approximately 60.07 and 59.65% of the total variance of all variables, respectively. FIT and AFIT were both statistically significantly different from zero (i.e., not include zero value). As a supplementary indicator of overall model fit, the Goodness-of-Fit Index (GFI) and Standardized Root Mean square Residual (SRMR) provide the proximity between sample covariance and covariance. GFI scores around 1 and SRMR values near 0 may be considered a decent fit. Results reported that the GFI value was very close to one, while the SRMR value was close to zero.

Table 7 displayed the estimates of path coefficients in the model specification, along with their standard errors and 95% confidence intervals. According to the results of the experiment, behavioral intention was statistically significant and positively impacted by performance expectancy ($H_1$ = 0.1003, SE = 0.1077, 95% CI = 0.1057-0.2826). Behavioral intention was also influenced by Trust positively ($H_5$ = 0.0968, SE = 0.1003, 95% CI = 0.0889-0.2913). Trust had a statistically significant and positive influence on Performance Expectancy ($H_6$ = 0.7612, SE = 0.0361, 95% CI = 0.6862-0.8294). In turn, Trust had a statistically significant and positive influence on Effort Expectancy ($H_7$ = 0.6795, SE = 0.05, 95% CI = 0.5619-0.7575) and Trust had a statistically significant and negative effect on Privacy Risk ($H_8$ = -0.1479, SE = 0.0604, 95% CI = -0.2541-0.0378). Moreover, Privacy risks had a statistically significant and negative influence on behavioral intention ($H_9$ = -0.0625, SE = 0.0624, 95% CI = 0.0678-0.1838). However, the three hypotheses $H_2$ (Effort Expectancy → Behavioral Intention) and $H_3$ (Social Influence → Behavioral Intention) and $H_4$ (Facilitating Conditions → Behavioral Intention) were not supported due to the presence of zero values.

**Table 2:** General profiles of the participants

| Variable | Item | Number | Percentage |
|---|---|---|---|
| Gender | Male | 36 | 16.07 |
| | Female | 188 | 83.93 |
| Age | 10-20 | 118 | 52.68 |
| | 21-30 | 61 | 27.23 |
| | 31-40 | 25 | 11.16 |
| | Over 40 | 20 | 8.93 |
| Living area | City | 64 | 28.57 |
| | Town | 43 | 19.20 |
| | Rural Area | 117 | 52.23 |
| Total | | 224 | 100.00 |

**Table 3:** Means and standard deviations of the personal traits and UTAUT's measures (N = 224)

| Construct | Item | Mean | SD |
|---|---|---|---|
| Performance expectancy | PE 1 | 4.3928 | 0.8799 |
| | PE 2 | 4.3303 | 0.8544 |
| | PE 3 | 4.2633 | 0.9436 |
| Effort expectancy | EE 1 | 4.4152 | 0.8139 |
| | EE 2 | 4.4017 | 0.7787 |
| | EE 3 | 4.4196 | 0.8088 |
| Social influence | SI 1 | 4.2455 | 0.9294 |
| | SI 2 | 4.3392 | 0.8869 |
| | SI 3 | 4.3705 | 0.8458 |
| Facilitating conditions | FC 1 | 4.5937 | 0.6880 |
| | FC 2 | 4.4866 | 0.8072 |
| | FC 3 | 3.5223 | 1.2245 |
| Privacy risk | PR 1 | 2.7768 | 1.3707 |
| | PR 2 | 2.7277 | 1.3469 |
| Trust | T 1 | 4.5045 | 0.7134 |
| | T 2 | 4.3705 | 0.8511 |
| | T 3 | 4.1429 | 0.9483 |
| Behavioral intention | BI 1 | 4.5714 | 0.7345 |
| | BI 2 | 4.4911 | 0.8398 |
| | BI 3 | 4.5357 | 0.7668 |

**Table 4:** Internal consistency and convergent validity

| Construct | Items | Dillon-goldstein's rho | AVE |
|---|---|---|---|
| Performance expectancy | 3 | 0.8711 | 0.9205 |
| Effort expectancy | 4 | 0.9194 | 0.9394 |
| Social influence | 3 | 0.7733 | 0.8721 |
| Facilitating conditions | 3 | 0.7394 | 0.8080 |
| Privacy risk | 2 | 0.8882 | 0.9470 |
| Trust | 3 | 0.8290 | 0.7453 |
| Behavioral intention | 3 | 0.8326 | 0.8997 |

**Table 5:** Estimates of loadings

|       | Estimate | Std. error | 95% CI LB | 95% CI UB |
|-------|----------|------------|-----------|-----------|
| PE 1  | 0.8719   | 0.0208     | 0.8373    | 0.9048    |
| PE 2  | 0.8903   | 0.0197     | 0.8542    | 0.9244    |
| PE 3  | 0.9110   | 0.0196     | 0.8688    | 0.9404    |
| EE 1  | 0.8361   | 0.0346     | 0.7562    | 0.8996    |
| EE 2  | 0.9158   | 0.0176     | 0.8704    | 0.9437    |
| EE 3  | 0.9181   | 0.0214     | 0.8636    | 0.9490    |
| EE 4  | 0.8944   | 0.0172     | 0.8656    | 0.9298    |
| SI 1  | 0.9045   | 0.0178     | 0.8640    | 0.9353    |
| SI 2  | 0.9405   | 0.0081     | 0.9237    | 0.9565    |
| SI 3  | 0.6310   | 0.0715     | 0.4694    | 0.7483    |
| FC 1  | 0.8545   | 0.0348     | 0.7912    | 0.9279    |
| FC 2  | 0.9034   | 0.0172     | 0.8693    | 0.9456    |
| FC 3  | 0.4987   | 0.0862     | 0.3228    | 0.6388    |
| T 1   | 0.8736   | 0.0244     | 0.8209    | 0.9148    |
| T 2   | 0.8863   | 0.0254     | 0.8293    | 0.9265    |
| T 3   | 0.8289   | 0.0291     | 0.7485    | 0.8747    |
| PR 1  | 0.9502   | 0.0094     | 0.9292    | 0.9642    |
| PR 2  | 0.9466   | 0.0089     | 0.9266    | 0.9586    |
| BI 1  | 0.8924   | 0.0339     | 0.8158    | 0.9436    |
| BI 2  | 0.8714   | 0.0295     | 0.8047    | 0.9258    |
| BI 3  | 0.8325   | 0.0426     | 0.7447    | 0.8999    |

**Table 6:** Model FIT

|                     | Estimate | SE     | 95% CI LB | 95% CI UB |
|---------------------|----------|--------|-----------|-----------|
| FIT                 | 0.6007   | 0.0140 | 0.5768    | 0.6291    |
| Adjusted FIT (AFIT) | 0.5965   | 0.0142 | 0.5724    | 0.6252    |
| GFI                 | 0.9930   | 0.0009 | 0.9916    | 0.9949    |
| SRMR                | 0.2419   | 0.0194 | 0.2171    | 0.2852    |

**Table 7:** Estimates of path coefficients

|                              | Estimates | Std. error | 95% CI LB | 95% CI UB |
|------------------------------|-----------|------------|-----------|-----------|
| PE → BI (H1)                 | 0.1003*   | 0.1077     | 0.1057    | 0.2826    |
| EE → BI (H2)                 | -0.0998   | 0.0862     | -0.2818   | 0.0581    |
| SI → BI (H3)                 | -0.0403   | 0.1109     | -0.2262   | 0.2101    |
| FC → BI (H4)                 | -0.0935   | 0.0914     | -0.3003   | 0.0750    |
| T → BI (H5)                  | 0.0968*   | 0.1003     | 0.0889    | 0.2913    |
| T → PE (H6)                  | 0.7612*   | 0.0361     | 0.6862    | 0.8294    |
| T → EE (H7)                  | 0.6795*   | 0.0500     | 0.5619    | 0.7575    |
| T → PR (H8)                  | -0.1479*  | 0.0604     | -0.2541   | -0.0378   |
| PR → BI (H9)                 | -0.0625*  | 0.0624     | 0.0678    | 0.1838    |

*Statistically significant at 0.05 level

## Discussion

It has been more than two years since the first outbreak of the COVID pandemic, the loss of life and property is unimaginable. Even though many nations have opened up and reverted to the new normal with the motto "living with COVID," (Wei *et al*., 2021) we are still seeing outbreaks of infections in some regions. Governments are still looking for ways to combat the infection (Fagherazzi *et al*., 2020; Kamran and Ali, 2021), from vaccination at the earliest and fullest dosage, to informing citizens and tracking sick individuals and prospective cases. In this combat, the COVID-19 tracing application is seen as viable, beneficial, and low-cost (Munzert *et al*., 2021). These tracing applications, however, have not realized their full potential (Walrave *et al*., 2021). The fact that the number of installers remains low in both industrialized and developing nations is indicative of this. There are several reasons leading to this consequence, including confidence in the network provider, worries about information security, and concerns about privacy (Williams *et al*., 2021). Furthermore, the haste with which applications are developed in a short amount of time will unavoidably result in system and software problems. For example, an application that crashes does not update information, or is incompatible with a wide range of devices contributes to lower user engagement.

According to the findings, performance expectations, trust, and privacy all have an impact on app usage intention. They both corroborated previous research and corresponded to real-world findings. However, social impact, effort expectation, and facilitating conditions were not shown to be statistically significant. It may suggest that the app is simple to use, comparable to a wide variety of devices and that family members, friends, or college are not predictors of the behavioral intention. This result is critical in carefully evaluating the aforementioned factors, particularly social influence. People in industrialized and Western nations live relatively autonomously and are less reliant on one another so social factors may not affect the app usage. However, in many countries, this relationship is getting tighter. As a result, social influence should be engaged and encouraged to contribute to app utilization.

The current study adds to the body of knowledge in two ways. First and foremost, the authors extended the Unified Theory of Acceptance and Use of Technology (UTAUT) theoretical model by highlighting the elements impacting behavioral intention to use the COVID-19 contact tracking app. Perhaps, the amount of variation explained by the extended UTAUT models (60.07%) was one of the most notable findings of the current research. Nonetheless, the UTAUT is a suitable paradigm for studying this type of technological behavior. To the best of our knowledge, this is one of the unique research to investigate the causal relationships between a collection of factors and behavioral intention to use a COVID tracker app, taking into consideration trust and privacy. As a result, the findings of this study add to the COVID-related area and user behavior literature by emphasizing the consequences of COVID-19 contact tracing app use in the real world. Second, by incorporating trust and privacy risk into the UTAUT model, we validated prior research, adding to the collection of knowledge on the topic.

Despite the contributions indicated above, the results are certain to be hampered by a variety of restrictions. These constraints, when paired with unanticipated discoveries, result in viable future study research directions. First, non-probability sampling was utilized in this research to ensure that respondents had applications loaded on their devices. This sampling approach, however, as commonly accepted in the literature, limits the generalizability of the findings beyond the sample characteristics given in this study. Thus, future work is encouraged to reexamine the proposed model with a random technique. Second, this study investigated the usage of the COVID-19 tracker over a short time. Given the rapid pace of technological advancement, the findings of the current research must be revisited as technology advances. Third, because the current work relied solely on two external factors as its theoretical framework, other components could not be evaluated. It is, therefore, necessary for researchers to examine additional potential factors

COVID-19 is still a current issue that many academics are interested in. Many proposals have been made to improve the efficacy of the COVID tracing application.

We do not attempt to include or provide all recommendations because some are only suited for a certain ethnic location. Instead, we only provide suggestions that are appropriate to the environment and culture in which this study was carried out. The first is for device software compatibility concerns. Application developers and network providers must collaborate closely to explore the many sorts of phones and the operating systems that operate on them. This information is critical in determining which platform to develop on to be compatible with many devices. Typically, the most recent libraries are only compatible with new devices and have issues with older devices. If most people continue to use outdated technology, the software will not reach the masses. The second step is to raise public awareness of the COVID pandemic and educate individuals on how to prevent and battle it. This is a tough challenge to solve since, in addition to combating the disease, we must simultaneously deal with inflation, food shortages, and job losses. People must still go out to work and communicate to exchange information and commodities. People who serve as hubs should be vaccinated first and then those who have recovered from COVID should be recruited to engage in socially important activities. The third concern is personal privacy. There should be an agreement in place, as well as explicit notification from software and service providers to users, outlining how the information gathered will be utilized. Previous research has revealed that many consumers are still apprehensive about utilizing applications that have access to their location. Although service providers may not always violate users' privacy, stolen or leaked data poses a significant danger that must be addressed.

## Conclusion

This study investigated the effects of variables influencing the intention to use the COVID-19 tracker. The extended Unified Theory of Acceptance and Use of Technology model was adapted to investigate user behavior toward experiencing the COVID-19 tracker application in terms of performance expectancy, effort expectancy, social influence, facilitating conditions, trust, and perceived risk. Based on the evidence from 224 participants, the findings revealed that the influence of performance expectation on behavioral intention was statistically significant and positive. In addition, Trust had a statistically significant and positive influence on behavioral intention. Trust also positively influenced Performance Expectancy. In turn, Trust had a positive influence on Effort Expectancy and Trust had a statistically significant and negative effect on Privacy Risk. Moreover, Privacy risks negatively impacted behavioral intention. However, hypotheses $H_2$ (Effort Expectancy $\rightarrow$ Behavioral Intention), $H_3$ (Social Influence $\rightarrow$ Behavioral Intention), and $H_4$ (Facilitating Conditions → Behavioral Intention) were not supported in the experiment. Overall, the proposed conceptual model explained 60.07% of the amount of variation, suggesting that software developers, service providers, and policymakers should consider performance expectations, trust, and privacy as viable factors to encourage citizens to use the app. Further study is called to investigate these non-significant correlations.

## Acknowledgment

The authors would like to thank all colleagues and students for taking part in the survey. We would like to thank Dr. Minh, Nguyen Hai for his valuable support.

## Author's Contributions

**Vinh T. Nguyen:** Participated in research conceptualization, data analysis, reporting, and manuscript preparation.

**Chuyen T. H. Nguyen:** Researched relevant studies, collected and pre-processes data, and revised the manuscript.

## Ethics

This article is unique and contains unpublished material. The comparing creator affirms that all different writers have perused and endorsed the composition what's more no moral issues are included.